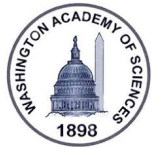

# PROSPECTIVE LAVA TUBES AT HELLAS PLANITIA

### LEVERAGING VOLCANIC FEATURES ON MARS TO
### PROVIDE CREWED MISSIONS PROTECTION FROM RADIATION

ANTONIO J. PARIS, EVAN T. DAVIES, LAURENCE TOGNETTI, & CARLY ZAHNISER
CENTER FOR PLANETARY SCIENCE

## ABSTRACT

Mars is currently at the center of intense scientific study aimed at potential human colonization. Consequently, there has been increased curiosity in the identification and study of lava tubes for information on the paleohydrological, geomorphological, geological, and potential biological history of Mars, including the prospect of present microbial life on the planet. Lava tubes, furthermore, could serve as *in–situ* habitats for upcoming crewed missions to Mars by providing protection from solar energetic particles, unpredictable high-energy cosmic radiation (i.e., gamma-ray bursts), bombardment of micrometeorites, exposure to dangerous perchlorates due to long-term dust storms, and extreme temperature fluctuations. The purpose of this investigation is to identify and study prospective lava tubes at *Hellas Planitia*, a plain located inside the large impact basin *Hellas* in the southern hemisphere of Mars, through the use of Earth analogue structures. The search for lava tubes at *Hellas Planitia* is primarily due to the low radiation environment at this particular location. Several studies by NASA spacecraft have measured radiation levels in this region at ~342 µSv/day, which is considerably less than other regions on the surface of Mars (~547 µSv/day). Notwithstanding, a radiation exposure of ~342 µSv/day is still sizably higher than what human beings in developed nations are annually exposed to on Earth. By analyzing orbital imagery from two cameras onboard NASA's Mars Reconnaissance Orbiter (MRO) – the High-Resolution Imaging Science Experiment (HiRISE) and the Context Camera (CTX) – the search for lava tubes was refined by identifying pit crater chains in the vicinity of *Hadriacus Mons*, an ancient low-relief volcanic mountain along the northeastern edge of *Hellas Planitia*. After surveying 1,500 images from MRO, this investigation has identified three candidate lava tubes in the vicinity of *Hadriacus Mons* as prospective sites for manned exploration. To complement this investigation, moreover, 30 *in-situ* radiation monitoring experiments have been conducted at analog lava tubes located at Mojave, CA, El Malpais, NM, and Flagstaff, AZ. On average, the total amount of solar radiation detected outside the analog lava tubes was approximately ~0.470 µSv/hr, while the average inside the analog lava tubes decreased by 82% to ~0.083 µSv/hr. We infer, therefore, that the candidate lava tubes identified southwest of *Hadriacus Mons* could be leveraged to decrease the radiation and to reduce the crew's exposure from ~342 µSv/day to ~61.64 µSv/day (a decrease of ~82%). This investigation, therefore, concluded that terrestrial lava tubes can be leveraged for radiation shielding and, accordingly, that the candidate lava tubes on Mars (as well as known lava tubes on the lunar surface) can serve as natural radiation shelters and habitats for a prospective crewed mission to the planet.

## THE RADIATION ENVIRONMENT ON MARS

Earth's magnetic field protects us from harmful solar and galactic cosmic radiation. Though astronauts in low-earth orbit are more exposed to radiation than humans on the ground, they are still protected by Earth's magnetosphere. Outside our magnetosphere, however, radiation is more problematic. Research studies of exposure to various strengths and doses of radiation provide strong evidence that degenerative diseases and/or cancer are to be expected from too much exposure to solar energetic particles and/or galactic cosmic rays.[1] Galactic cosmic radiation contains highly ionizing heavy ions that have large penetration power in tissue and that may produce extremely large doses, leading to early radiation sickness or death if adequate shelter is not provided.[2] During the Apollo program, for illustration, astronauts on the moon reported headaches, reported seeing flashes of light, and experienced painful cataracts.[3] These symptoms, known as Cosmic Ray Visual Phenomena, were due to radiation from cosmic rays interacting with matter and deposing its energy directly in the eyes of the astronaut.[4] The Apollo missions, however, were comparatively short, and they cannot be likened to a multi-year presence on the surface of Mars.

Approximately 4.2 billion years ago, due to either rapid cooling in its central core or a massive impact from an asteroid or comet, the magnetic dynamo effect on Mars stopped, and its magnetosphere weakened dramatically.[5] As a result, over the course of the next 500 million years, the Martian atmosphere was gradually stripped away by the solar wind. Between the loss of its magnetic field and its atmosphere, the surface of Mars is now exposed to much higher levels of solar and cosmic radiation than Earth. In addition to regular exposure to solar energetic particles and galactic cosmic rays, the planet receives intermittent harmful blasts that occur from strong solar flares, as well as the bombardment of meteors.[4] A crewed mission to the surface of Mars, consequently, will introduce the crew and its critical life-support equipment to an environment outside a much-needed magnetosphere. They will be at risk of absorbing sudden fatal radiation doses, as well as assuredly suffering cellular and DNA damage from chronic high background radiation, which will lead to cancer. Additionally, there is a risk of an unpredictable cosmic ray burst or a meteor shower capable of critically damaging the crew's life-support equipment.[1]

Numerous investigations by NASA have established that the surface of Mars receives a varying amount of radiation. The difference is largely due to where the solar wind interacts with electrically charged particles in the Martian upper atmosphere and the particular topographic elevation on the planet. The Mars Atmosphere and Volatile Evolution (MAVEN) spacecraft, for illustration, detected the interaction of the solar wind with electrically charged particles of the Martian atmosphere.[6] Atoms in the upper atmosphere became electrically charged ions after being energized by solar and cosmic radiation. Because they are electrically charged, these ions interact with the magnetic and electric forces of the solar wind, a thin stream of electrically conducting gas ejected from the surface of the Sun.[4] As explained in Figure 1, the lines represent the paths of individual ions, and the colors represent their energy, E, measured in electron volts (eV). The MAVEN investigation concluded that the northernmost polar plume contains the most energetic ions, ranging as high as ~20,000 E (eV) – with the surface of the planet's northern polar region receiving approximately ~10,000 to ~1,000 E (eV), the southern polar region receiving ~1,000 to ~100 E (eV), and the mid-latitudes receiving the lowest at ~300 to ~10 E (eV).[7]

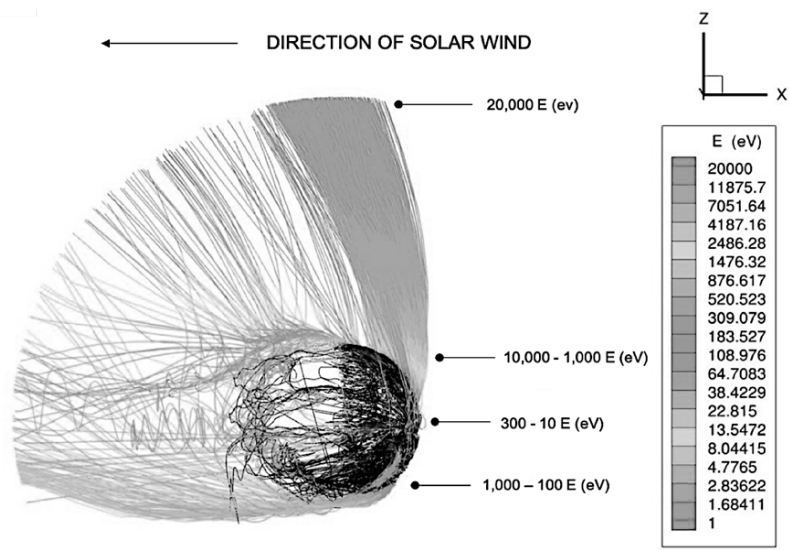

**Figure 1:** MAVEN Investigation of the Solar Wind on Mars.
(Source: NASA)

Similarly, the NASA Mars Odyssey spacecraft was equipped with a special instrument called the Martian Radiation Experiment (MARIE), which was also designed to measure the radiation environment on Mars.[8] Over the course of 18 months, MARIE detected ongoing levels of radiation that were considerably higher than what astronauts experience on the International Space Station (~200 µSv/day).[5] At high topographic elevations, MARIE detected radiation levels at ~547 µSv/day, which equates to ~200,000 µSv/year (Figure 2). Although there are no immediate symptoms while being exposed to ~547 µSv/day, exposure to these levels of radiation for one year is four times the maximum annual allowable exposure for U.S. radiation workers (50,000 µSv/year), which can cause radiation sickness.[9] In contrast, the radiation levels at lower elevations were measure at ~273 µSv/day, because those areas have more atmosphere above them to block out a considerable amount of the radiation.[10] For comparison, humans in developed nations are exposed to, on average, ~6,200 µSv/year.[11]

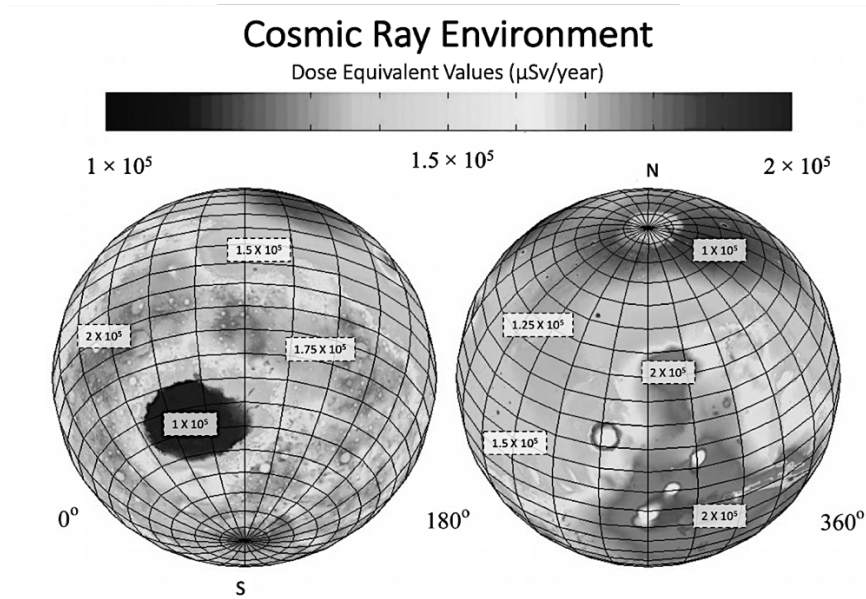

**Figure 2:** MARIE Investigation of the Cosmic Radiation Environment on Mars.
(Source: NASA)

In September 2017, strong energetic particles from a coronal mass ejection were detected both in Mars orbit and on the surface (Figure 3).[12] In orbit, MAVEN detected energetic particles as high as 220 million E (eV), which produced radiation levels on the surface more than double any previously measured by Curiosity's Radiation Assessment Detector (RAD).[13] RAD is an energetic particle detector capable of measuring all charged particles that contribute to the radiation health risks that future crewed missions to Mars will face.[14] At the time of the coronal mass ejection, Curiosity was positioned inside Gale Crater at an elevation of ~4,500 m. As cited previously, a lower elevation on Mars allows for additional atmosphere above the rover, which aids in blocking out a considerable amount of radiation. At Gale Crater, the normal background radiation measured by RAD prior to the coronal mass ejection was ~220-270 µSv/day, which is similar to what MARIE detected (~273 µSv/day). However, during the coronal mass ejection, RAD detected radiation levels at ~600 µSv/day – over twice the normal background radiation at Gale Crater (Figure 3).[12] While studies have shown that the human body can withstand a single dose of up to 2,000,000 µSv without permanent damage,[4] prolonged exposure to strong energetic particles will lead to health complications – including acute radiation sickness (nausea, vomiting, weakness, headaches, purpura, hemorrhage, infections, diarrhea, leukopenia), genetic damage, and possibly death. Moreover, exposure to more than 5,000,000 µSv would kill half of those exposed within a month, and 10,000,000 µSv would be fatal within days.[15]

The data from MAVEN, MARIE, and RAD show that the solar wind and other violent solar activity, such as solar flares and coronal mass ejections, continue to strip away the Martian atmosphere – but more importantly, they make the northern latitudes less desirable for human exploration. The data indicate, moreover, that a crewed mission should more prudently be placed along the mid-to-southern latitude range – where there is minimized ionic activity from the solar wind (i.e., *Hellas Planitia*). Furthermore, the data suggest that the crew should be placed at lower elevations along the surface of Mars (again, *Hellas Planitia*), with its thicker atmosphere, which will help to block out the radiation by a factor of two. Additionally, a crewed mission on Mars should be effectuated during a solar minimum (the period of least solar activity in the 11-year solar cycle of the Sun) to ensure that exposure to radiation due to a solar flare and/or coronal mass ejection is minimized.[16]

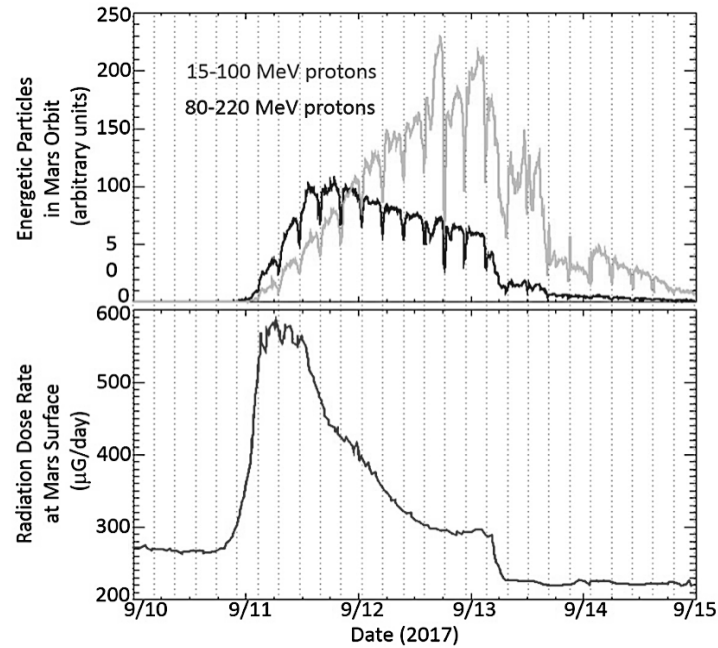

**Figure 3:** Radiation data collected by MAVEN and RAD during the September 2017 solar storm.
(Source: NASA)

## ANCIENT VOLCANISM ON MARS

Volcanic activity has played an extensive role in the geologic development of Mars. Planetary scientists have acknowledged since the Mariner 9 mission in 1972 that volcanic topographies cover great portions of the Martian surface. These geological features include widespread lava flows, lava plains, and the largest known volcanic peaks in the solar system, including the highest, Olympus Mons.[17] Volcanic features on Mars range in age from the Noachian (> 3.7 billion years) to the late Amazonian (< 500 million years) eras, demonstrating that Mars was volcanically active early in its history.[18] Mars, moreover, is a differentiated terrestrial planet that formed from analogous chondritic materials.[19] Many of the same magmatic processes that occurred on Earth, for comparison, also took place on Mars. The two planets are sufficiently comparable compositionally that similar names can be applied to their igneous rocks and minerals.[20] The shape and size of volcanoes on Mars, though, were largely dependent on a set of environmental conditions and properties dissimilar to Earth. Lesser atmospheric pressure altered the scattering of ejected material, a higher eruption rate allowed for the lava on the surface to pile up higher, and lower gravity facilitated wider dispersion.[21] Consequently, volcanoes and lava tubes on Mars are more than twice as large as their terrestrial analogues.

## THE FORMATION OF LAVA TUBES

Lava is molten rock that is ejected by a volcano during an eruption. The molten rock on Mars was formed in the interior of the planet in high temperatures produced by geothermal energy. When the lava first erupts from a volcanic vent, it is in a liquid state, typically at temperatures ranging from 1,292° to 2,192°F.[22] A lava flow, in contrast, is an outburst of lava that moves during a non-explosive effusive eruption. When the lava flow stops moving, it cools and hardens to form igneous rock. While lava can be up to 100,000 times more viscous than water, it can flow for a large distance before abating and solidifying.[23] Lava flows can occasionally form lava tubes – natural subsurface caves or caverns – which appear to form because of rapid lava flow. Lava tubes, which are usually made from extremely fluid pahoehoe lava, typically form when the exterior surface of lava channels cools more rapidly and forms a strong crust over the subsurface lava.[24] The lava flow ultimately stops and drains out of the tube, leaving a conduit-shaped void located several feet underneath the surface (Figure 4). As we discussed above, as gravity on Mars is 37% of that on Earth, lava flow, and subsequently lava tubes on Mars, are generally much larger than those found on Earth.[25]

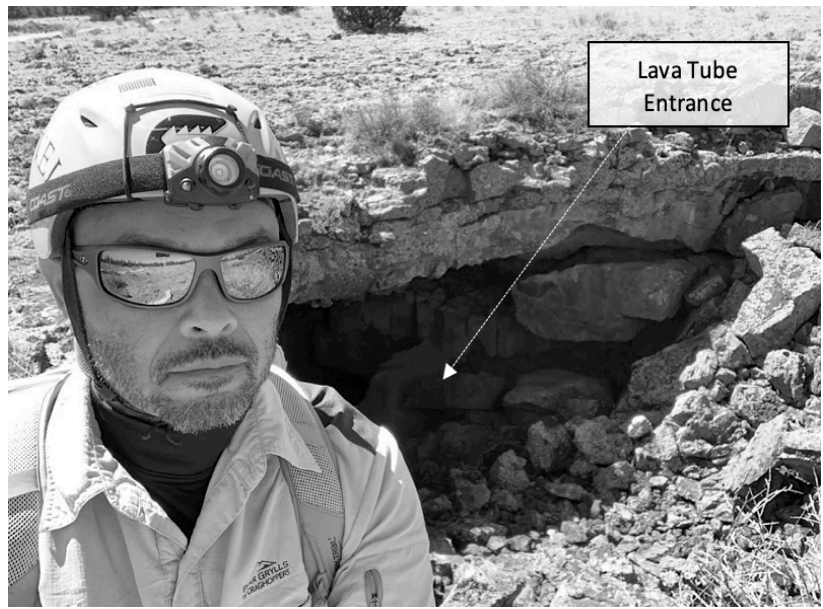

**Figure 4**: Prof. Antonio Paris, the Principal Investigator, conducting research at the El Calderon lava tubes in El Mapais, New Mexico.

## IDENTIFYNG PIT CRATER CHAINS AND LAVA TUBES

Lava tubes below the surface of Mars, as well as those below the lunar surface, can be identified by locating pit craters in the vicinity of known ancient lava flows. A pit crater is a circular or elliptical depression shaped by the collapsing or sinking of the surface lying above a void or hollow cavity, rather than by the eruption of a lava vent or volcano. Pit craters generally lack a raised rim, uplifting, ejecta blankets associated with impact craters, or radial patterns of lava flows discharging downslope from volcanic calderas.[26] There are generally two types of pit craters – atypical and bowl-shaped. Atypical pit craters exhibit a distinctive set of morphologies and characteristics that set them apart from the commonly observed bowl-shaped pit craters. Instead of bowls, the interior of atypical pit craters is cylindrical or bell-shaped with vertical to overhanging walls that extend down to their floors without forming talus slopes.[27] In some instances, atypical pit craters show evidence of vents, fissures, and caverns/caves that could be intact.

Pit craters have been discovered on Earth, Mars, as well as on the various moons in our solar system (including our Moon).[28] They are generally located in a series of ranged or offset chains; and in these instances, they are referred to as pit crater chains (Figure 5). When adjacent walls between pits in a pit crater chain collapse, they become troughs.[29] A commonly invoked hypothesis to explain these troughs is that they are collapsed lava tubes, essentially tunnels formed underground by rivers of lava.[30]

Through the use of spacecraft imagery, lava tubes on Earth, Mars and the Moon can be visually detected in two ways. The first method is by recognizing troughs or rilles, which, as mentioned previously, are believed to be the remains of collapsed lava tubes. The second method is by pinpointing "skylights" – dark, nearly rounded features that are hypothesized as entrances to lava tubes.[31] At this point, light from the Sun enters into the permanent darkness of the lava tube from above, forming a skylight.[20] Many lava tubes on Earth, for instance, have been identified through the discovery of skylights (Figure 5). As is the case on Earth, access to uncollapsed sections of lava tubes on Mars is available by entering through a skylight, at the end of the trough or rille where a cave/cavern could lead into the lava tube, or by drilling or blasting through the roof of a lava tube.[32] The Lunar Reconnaissance Orbiter, moreover, has imaged over 200 pits craters on the lunar surface. Many of these lunar pits craters show skylights into subsurface voids or caverns, ranging in diameter from about 5m to more than 900m, although some of these are likely to be post-flow features rather than volcanic skylights.[33] Moreover, there is observational evidence from orbiting spacecraft to infer there are lava tubes along the *Marius Hills*, *Hadley Rille* and *Mare Serenitatis* regions of the Moon. [34] While the lunar surface varies in temperature from -180 C to +100 C, the interior of these lunar lava tubes could remain at a constant -20 C. Therefore, lava tubes on the Moon (as well as on Mars), once sealed off, could be warmed up and pressurized with a breathable atmosphere.[35]

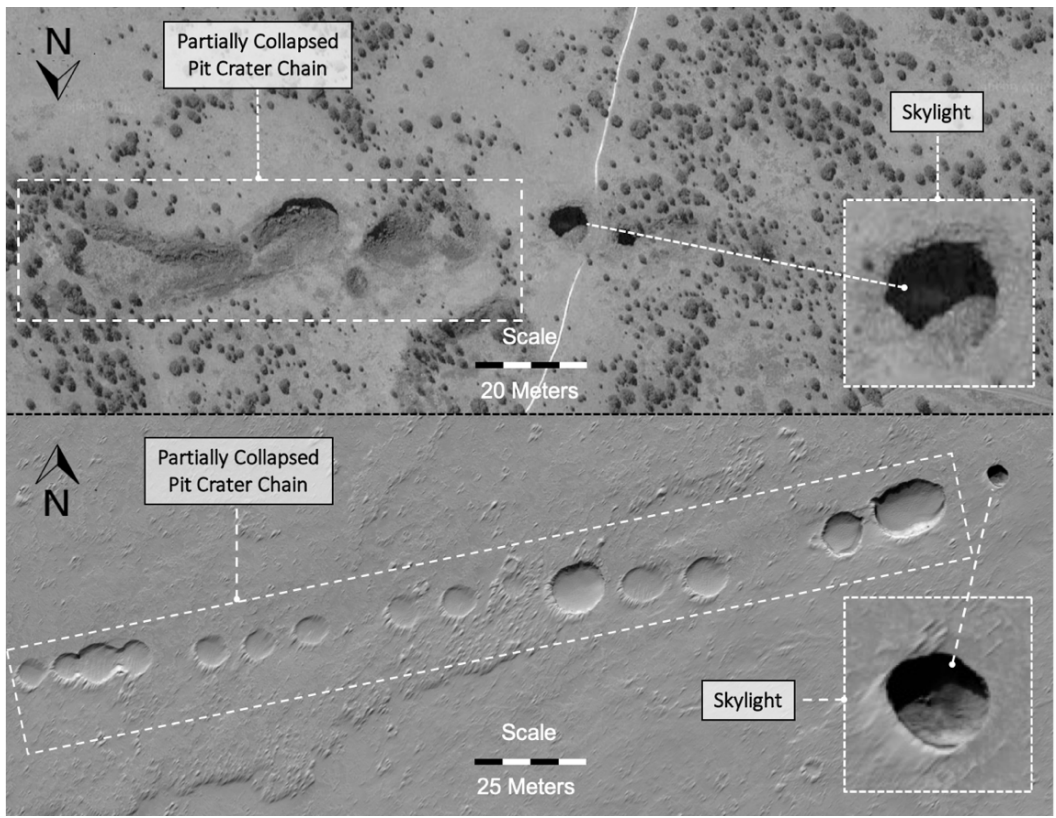

**Figure 5:** Similarities between a pit crater chain and pit crater "skylight" at El Mapais, NM (top image) and South of Arsia Mons on Mars (bottom image). (Source: NASA, Processing: Center for Planetary Science)

## HELLAS PLANITIA:
## PROSPECTIVE SITE FOR MANNED EXPLORATION

*Hellas Planitia* is a large meteorite impact basin located at 42° 42' S and 70° 00' E, in the southern hemisphere of Mars, in the *Hellas* and *Noachis* quadrangle.[36] The impact basin is thought to have been formed during the Late Heavy Bombardment period of the Solar System, approximately 4.1 to 3.8 billion years ago, when a large asteroid hit the surface. It is 2,300 km in diameter, and it is one of the largest known impact craters in the solar system (Figure 6).[37] Measurements made by the Mars Orbiter Laser Altimeter (MOLA) instrument onboard the Mars Global Surveyor spacecraft indicate that the basin floor is ~7,152 m deep – making *Hellas Planitia* one of the bottom-most geographic elevations on the planet.[38] At this depth, consequently, *Hellas Planitia* has one the lowest ongoing radiation levels on Mars.

According to data from NASA's MARIE experiment, radiation levels at *Hellas Planitia* have been measured at ~1 x $10^5$ µSv/year, as opposed to the higher topographic elevations on Mars, which were measured at ~2 x $10^5$ µSv/year (Figure 6).[8] The depth of the basin, moreover, is below the standard topographic datum of Mars, which explains the atmospheric pressure at the bottom of the basin: 12.4 mbar during the northern summer. This is 103% higher than the pressure at the topographical datum (6.1 mbar) and above the triple point of water, indicating that a liquid phase might be present under certain conditions of temperature, pressure, and dissolved salt content.[39] These conditions could provide the crew on Mars the opportunity to identify and study lava tubes in this region for information on the geological, paleohydrological, overall geomorphological, and potential biological history of the planet. While there is as of yet no evidence for this, an exciting notion is that, as with extremophile environments on Earth, these locations may serve as habitats for possible microbial life on Mars.[40]

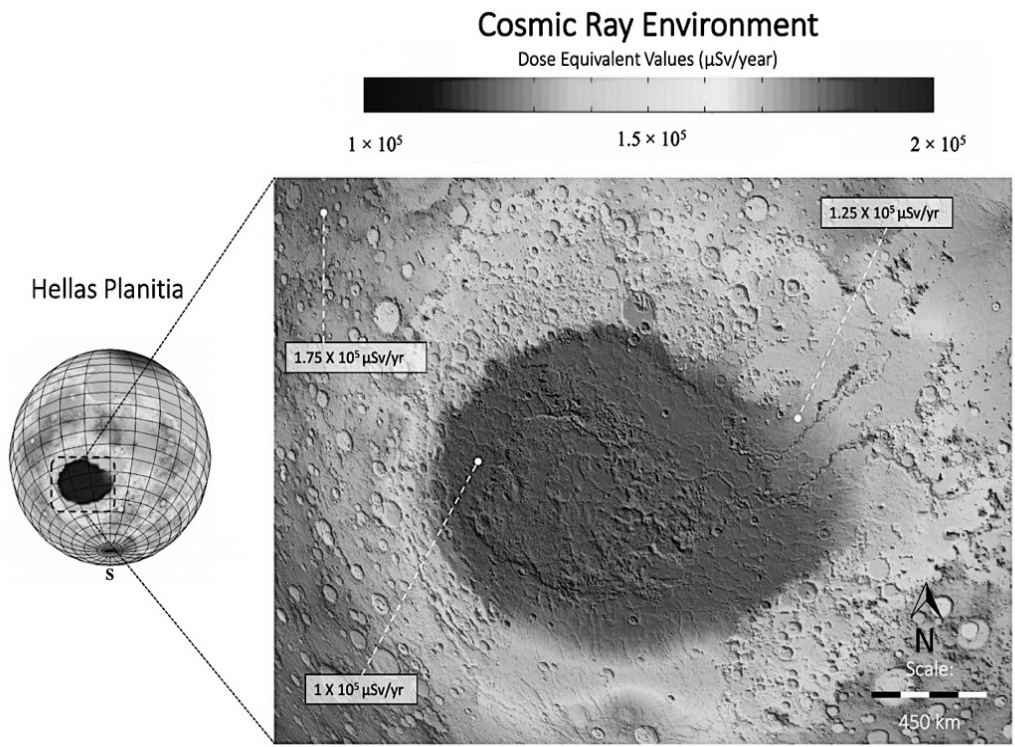

**Figure 6:** An expanded view of the radiation environment at Hellas Planitia.
(Source: Center for Planetary Science)

# HADRIACUS MONS REGION:
## PROSPECTIVE LAVA TUBES FOR MANNED EXPLORATION

*Hadriacus Mons* is an ancient, low-relief volcanic mountain located along the northeastern edge of *Hellas* (Figure 7). The volcano has a diameter of 450 kilometers, and its features differ from other volcanos such as Olympus Mons and the others in Tharsis and Elysium. *Hadriacus*'s low slopes, wide structure, and heavily scored flanks suggest that it is made of eroded materials.[41] Planetary scientists classify this volcanic debris as pyroclastic, from the Greek meaning "fire-broken." The large extent of volcanic deposits and the caldera size, moreover, leads some planetary scientists to infer that these geological topographies were the result of an explosive event caused by a contact between erupting magma and groundwater.[42]

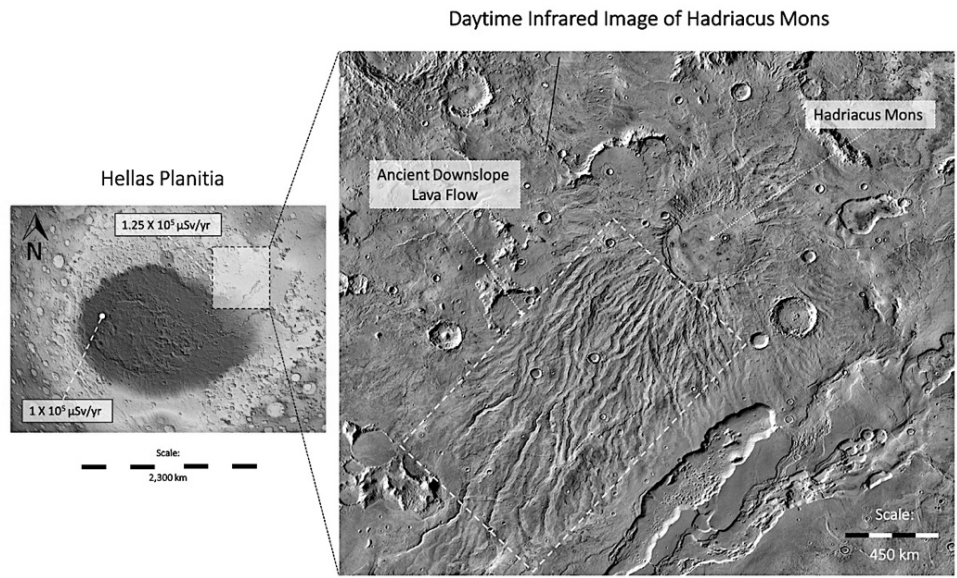

**Figure 7:** An expanded view of Hadriacus Mons along the edge of Hellas Planitia
(Source: NASA and Center for Planetary Science)

The low radiation levels along the lava flows in this region that extends from *Hadriacus Mons* into *Hellas Planitia* make the location a prime candidate for human exploration. According to data from NASA's MARIE experiment, radiation levels in this region have been measured at ~1.25 x $10^5$ μSv/year (~342.46 μSv/day), which, as previously mentioned, is considerably lower than the higher topographic elevations on Mars, which were measured at ~2 x $10^5$ μSv/year. Notwithstanding, a radiation exposure of 1.25 x $10^5$ μSv/year is still sizably higher than what human beings are annually exposed to on Earth. The lava tubes near *Hadriacus Mons*, consequently, could be used as natural radiation shelters and habitats for a crewed mission to the planet. These natural caverns have roofs estimated to be tens of meters thick, which would provide the crew protection from not only exposure to too much radiation, but also the bombardment of micrometeorites, exposure to dangerous soil perchlorates due to long-term dust storms, and extreme temperature fluctuations.[43] Moreover, although the exact conditions of the interior of Martian lava tubes will remain unknown until they are actually explored, planetary scientist are of the consensus that they represent prime locations for direct observation of pristine Martian bedrock, where keys critical to understanding the natural history of this planet will be found.[33]

The NASA HiRISE and CTX data used in this study were available through NASA's Planetary Data System (PDS). MRO CTX observes Mars's surface at ~6 m/pixel in swaths of 30 km across and up to ~160 km in length. MRO also observes the Martian surface earlier in the day; thus, more pit craters can be observed with partially sunlit floors.[32] A partially sunlit floor allows planetary scientists to identify specific pit crater characteristics such as skylights, individual boulders, dusty, or rocky surface textures, overhanging rims, wall and floor morphologies, and bedrock stratification.[32]

An analysis of 1,500 HiRISE and CTX images of the *Hellas Planitia* basin uncovered several volcanic features southwest of *Hadriacus Mons*, which were subsequently identified as candidate lava tubes (Figure 8). The first candidate lava tube identified in this study is located southwest of *Hadriacus Mons* in the *Dao Vallis* region at latitude -36.961° and longitude 87.841°E (MRO CTX Catalog: J11_049132_1423_XN_37S272W). The ~4,500-meter collapsed lava pit (Figure 8a), which is positioned between two partially collapsed sinuous pit crater chains, depicts two surface areas that have not collapsed (Figure 8a and 8b). Imagery analysis indicates the area of the northern uncollapsed surface is ~600 meters long and ~300 meters wide (Figure 8a), and that the area of the southern uncollapsed surface is ~900 meters long and ~600 meters wide (Figure 8b). Further analysis of the ~4,500-meter collapsed lava pit with its contrast range limited to low-end radiance values indicate that the pit crater walls encompassing both uncollapsed areas are structurally intact – ruling out a lava bridge – indicating that a lava tube below the surface is plausible, since these parts of the pit crater chain have not collapsed.

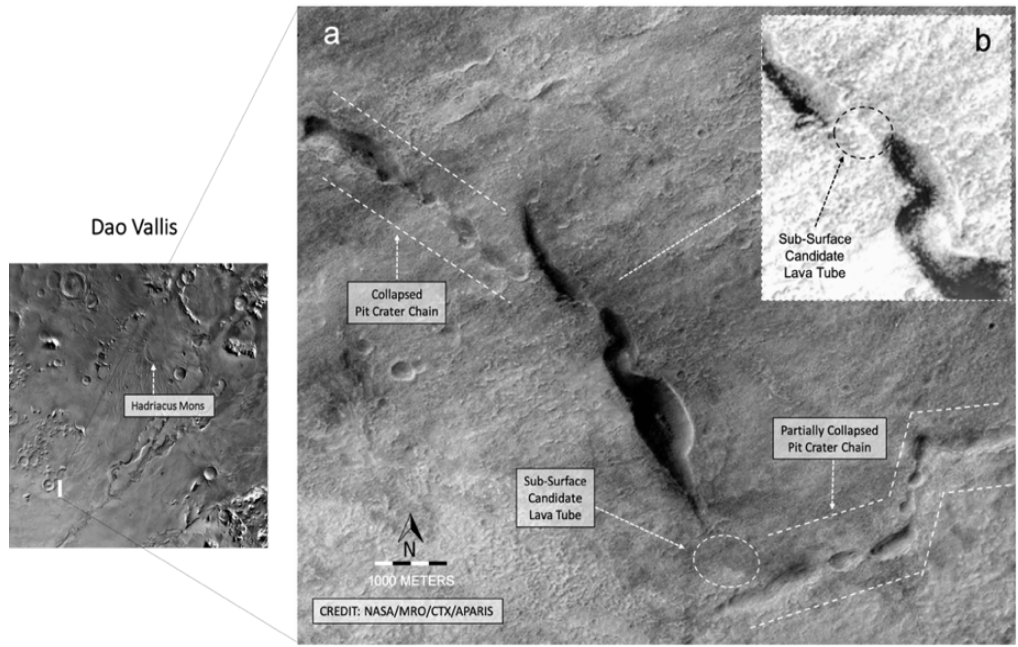

**Figure 8:** Candidate lava tubes south of Hadriacus Mons. (Credit: Center for Planetary Science)

The second candidate lava tube identified in this study is located south of *Hadriacus Mons* in the *Dao* and *Niger* valleys region at latitude -33.256° and longitude 93.980°E (MRO CTX Catalog: CTX: P13_006237_1475_XN_32S266W). An analysis of the CTX imagery depicts an atypical pit crater with a diameter of ~700 meters positioned between two sinuous chains of partially collapsed elliptical bowl-shaped pit craters (Figure 9a). The same atypical pit crater with its contrast range limited to low-end radiance values depicts the presence of a possible overhanging rim or shoulder along the northern section, which could be the entrance to a lava tube (Figure 9b). As mentioned previously, the leading accepted hypothesis is that since the pit crater chains East and West of this atypical pit crater have not fully collapsed, then a lava tube could exist below the surface.

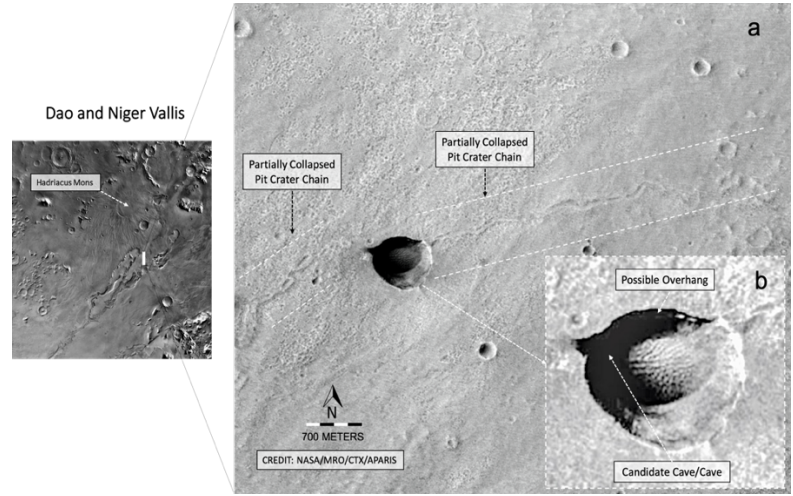

**Figure 9:** Candidate lava tubes south of Hadriacus Mons. (Credit: Center for Planetary Science)

The third candidate lava tube identified in this study is located southwest of *Hadriacus Mons* in the *Dao Vallis* region at latitude -36.870° and longitude 89.498°E (MRO CTX Catalog: F03_036987_1432_XN_36S270W). The ~25-meter collapsed pit, which appears to indicate the entrance to a candidate cavern and/or cave, is preceded by a partially collapsed trough (Figure 10a). The collapsed trough is sinuous, and, as previously cited, it could be associated with the presence of a lava tube below the surface. The same collapsed pit with its contrast range limited to low-end radiance values depicts the presence of boulders that more than likely arose from the collapse of the western wall section, which could be the entrance to a lava tube (Figure 10b). The region of the candidate lava tube, moreover, is characterized by numerous sinuous collapsed pit crater chains ranging in size – providing further evidence that one or more lava tubes could exist below the surface. Interestingly, this proposed lava tube is analogous to Giant Ice Cave in El Mapais, NM (Figure 11), which is also characterized by a partially collapsed sinuous trough that leads into the entrance of a lava tube.

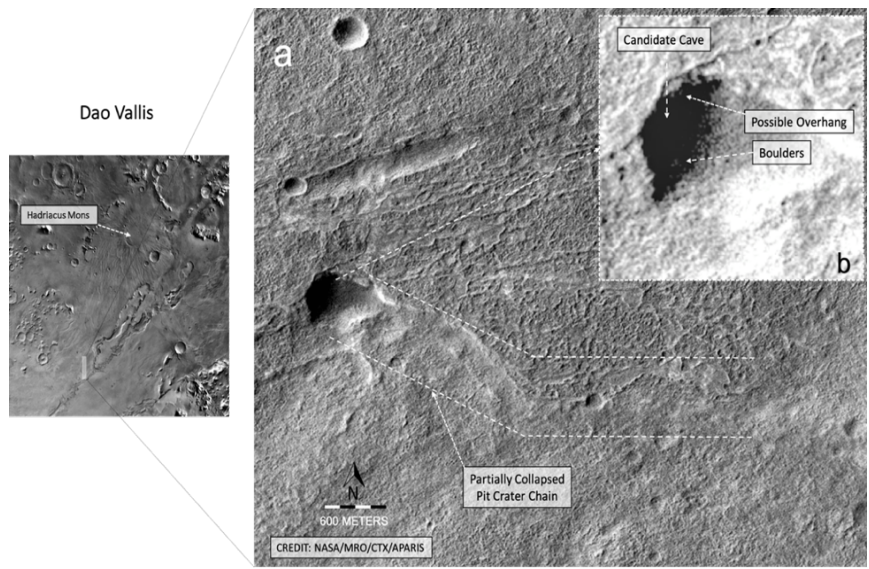

**Figure 10:** Candidate lava tubes south of Hadriacus Mons. (Credit: Center for Planetary Science)

# RADIATION EXPERIMENTS AT TERRESTRIAL ANALOG LAVA TUBES

To ascertain whether lava tubes on Mars (or correspondingly the lunar surface) could be used by surface crews to reduce exposure from solar or cosmic radiation, we conducted 30 analog radiation experiments. The experiments were conducted during solar noon at terrestrial lava tubes (Figure 11) located at Mojave, CA (Aiken), El Mapais, NM (Big Skylight, Giant Ice Cave, and Junction Cave) and Flagstaff, AZ. (Lava River Cave). Because radiation from the sun strikes the surface of Earth at different angles, conducting the experiments at solar noon (when the sun is directly overhead) allowed the surface area where the experiments were conducted to receive the most electromagnetic energy from the sun possible.[44] Furthermore, a sun at 90° above the surface of each lava tube prevented the radiation from becoming too scattered and diffused.

The Geiger counter for this experiment was equipped with a halogen-quenched, long thin cylindrical Geiger-Mueller tube capable of detecting Gamma and X-rays (ionizing radiation) down to 10 keV thru the window, 40 keV minimum through the case, and with a Gamma sensitivity of 10,000 μSv/hr. Although Earth's atmosphere and magnetic field protect us from most ionizing radiation from the Sun, streams of ionizing radiation still reach the surface of Earth. Occasionally, during a solar particle event, larger amounts of ionizing radiation strike the surface of the Earth, which can also be detected by the Geiger counter.

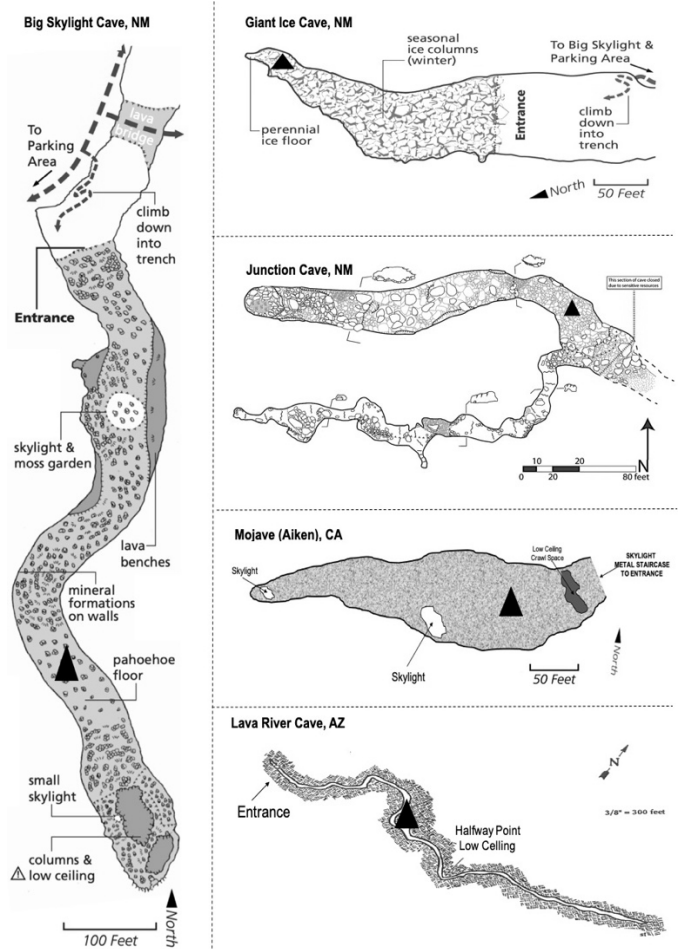

**Figure 11:** Lava tube maps and the interior locations (black triangle) where each radiation experiment was conducted.
(Map Sources: Mojave lava tube sketched by Prof. Antonio Paris, others National Park Service)

Preceding each experiment, the device was turned on for a minimum of one hour to establish a baseline reading. The experiment was then conducted in two parts – measuring the amount of solar radiation outside the lava tube and then comparing it with the amount of solar radiation measured inside the lava tube. In total, six one-hour readings were completed for each lava tube (three exterior and three interior) for a total of 30 observation hours (Table 1). On average, the total amount of radiation detected during solar noon on the exterior of all five lava tubes was ~0.471 μSv/hr ($V_1$) while the average radiation detected inside the lava tubes was ~0.083 μSv/hr ($V_2$). Furthermore, during the experiments, we observed a significant drop in temperature (Table 1) inside the terrestrial lava tubes, which likewise would be the case for lava tubes on Mars.

| Analog Site | Avg Surface Reading uSv/hr | Average Interior Reading uSv/hr | Avg Surface Temperature - C | Average Interior Temperature - C |
|---|---|---|---|---|
| Mojave Aiken - California | 0.468 | 0.088 | 12 | 9 |
| El Malpais - Big Skylight - New Mexico | 0.501 | 0.062 | 20 | 13 |
| El Malpais - Giant Ice Cave - New Mexico | 0.434 | 0.098 | 18 | 10 |
| El Malpais - Junction Cave - New Mexico | 0.458 | 0.078 | 17 | 11 |
| Lava River Cave - Arizona | 0.492 | 0.089 | 25 | 12 |
| **Average** | **0.4706** | **0.083** | **18.4** | **11** |

**Table 1:** Average amount of radiation (μSv/hr) measured at five terrestrial lava tubes.
(Source: The Center for Planetary Science)

When we applied the percentage change equation (Figure 12), the results of the radiation experiment showed that, on average, the amount of radiation ($A_l$) in the interior of the analog lava tubes decreased by an average of 82%. We can infer, therefore, that the candidate lava tubes identified southwest of *Hadriacus Mons* could be leveraged to decrease the radiation by ~82% and, accordingly, to reduce the crew's exposure from ~342.46 μSv/day to ~61.64 μSv/day.

$$A_l \; = \; \frac{\left| \Delta V \right|}{\left| V_1 \right|} \; \text{x} \; 100 \quad = \quad A_l \; = \; \frac{\left| V_2 - V_1 \right|}{\left| V_1 \right|} \; \text{x} \; 100$$

$$A_l \; = \; \frac{\left| 0.083 \; \mu Sv/hr \, - \, 0.470 \; \mu Sv/hr \; \right|}{\left| 0.470 \; \mu Sv/hr \; \right|} \; \text{x} \; 100 \; =$$

$$= 82.34\% \; (decrease \; in \; radiation)$$

**Figure 12:** Percentage change equation for radiation reduction inside analog lava tubes.

## CONCLUSION

An analysis of HiRISE and CTX imagery of the *Hellas Planitia* basin, specifically southwest of *Hadriacus Mons,* identified pit crater chains consistent with known lava tube morphology. Although the internal structural conditions of the candidate lava tubes remain largely unknown, a close examination of the satellite surface imagery suggests that sections of the pit crater chains have not collapsed, and therefore that lava tubes below the surface could be internally intact. Furthermore, the candidate lava tubes identified in this investigation are positioned in a region of Mars that regularly has lower radiation exposure than other regions on the planet. Though a background radiation environment of ~342.46 µSv/day is still significantly high, the terrestrial analog experiments conducted at Mojave, CA, El Malpais, NM, and Flagstaff, AZ concluded the lava tubes southwest of *Hadriacus Mons* could reduce the crew's exposure to radiation to ~61.64 µSv/day. The results of this investigation, therefore, indicate that the proposed lava tubes southwest of *Hadriacus Mons* can and should be utilized to serve as natural shelters for a crewed mission to the planet. These natural caverns would provide the crew protection from excessive radiation exposure, shelter them from the bombardment of micrometeorites, reduce their exposure to hazardous perchlorates in the Martian regolith, and provide them a degree of protection from extreme temperature fluctuations. The candidate lava tubes, moreover, can serve as important locations for direct observation and study of Martian geology and geomorphology, as well as potentially uncovering any evidence for the development of microbial life early in the natural history of Mars.

## BIOGRAPHY

Antonio Paris is the Chief Scientist at the Center for Planetary Science, a former Assistant Professor of Astronomy and Astrophysics at St. Petersburg College, FL, and a graduate of the NASA Mars Education Program at the Mars Space Flight Center, Arizona State University. He is the author of *Mars: Your Personal 3D Journey to the Red Planet*, and his latest peer-reviewed publications include *"El Bahr: A Prospective Impact Crater in Egypt"* – an investigation addressing the discovery of an unidentified crater located south of the Sahara Desert between Qaret Had El Bahr and Qaret El Allafa, Egypt. Prof. Paris is a professional member of the Washington Academy of Sciences and the American Astronomical Society, and he has appeared on the Science Channel, the Discovery Channel, and the National Geographic Channel.

## FIELD RESEARCH CONTRIBUTERS

Evan Davies is an anthropologist, science writer and a practicing physician assistant in emergency medicine. He is the author of *Emigrating Beyond Earth: Human Adaptation and Space Exploration*, and he is a fellow of both the Royal Geographical Society and the Explorers Club.

Laurence Tognetti recently graduated from Arizona State University with a Master's Degree in Geological Sciences, with thesis research focusing on geomorphological processes of the Martian surface. As a Field Researcher for the Center for Planetary Science, he collected and analyzed ionizing radiation data and geological specimens *in situ* for the analog experiments of this investigation.

Carly E. Zahniser recently graduated from Eckerd College with a Bachelor of Science in Environmental Science. As a Field Researcher for the Center for Planetary Science, she collected and analyzed ionizing radiation data and geological specimens *in situ* for the analog experiments of this investigation.

# REFERENCES

[1] National Aeronautics and Space Administration. Why Space Radiation Matters (available at https://www.nasa.gov/analogs/nsrl/why-space-radiation-matters/ accessed May 19, 2019).

[2] National Aeronautics and Space Administration. Space Radiation Cancer Risk Projections for Exploration Missions, JSC-29295 (available at https://pdfs.semanticscholar.org/1edf/1764e7d4cb9648e0e759ea235cac2db755ce.pdf).

[3] Hecht, Selig; Shlaer, Simon; Pirenne, Maurice Henri. (July 1942). "Energy, Quanta, and Vision." Journal of General Physiology. 25 (6): 819–840.

[4] National Aeronautics and Space Administration. Apollo Flight Journal (available at http://history.nasa.gov/afj accessed February 19, 2019).

[5] Williams, Matt. (November 21, 2016). "How Bad is the Radiation on Mars?" Universe Today (available at https://www.universetoday.com/14979/mars-radiation1/).

[6] National Aeronautics and Space Administration. Solar Energetic Particle Instrument for MAVEN Spacecraft (available at https://mars.nasa.gov/resources/5177/solar-energetic-particle-instrument-for-maven-spacecraft/).

[7] National Aeronautics and Space Administration. MAVEN Results Find Mars Behaving Like a Rock Star (available at https://www.nasa.gov/feature/goddard/rock-star-mars).

[8] National Aeronautics and Space Administration. MARIE Technical Features (available at https://mars.nasa.gov/odyssey/mission/instruments/marie/).

[9] Matson, John. (March 16, 2011). "Fast Facts about Radiation from the Fukushima Daiichi Nuclear Reactors." Scientific American (available at https://www.scientificamerican.com/article/japan-nuclear-fallout/).

[10] National Aeronautics and Space Administration. Estimated Radiation Dosage on Mars (available at https://www.jpl.nasa.gov/spaceimages/details.php?id=PIA03480).

[11] U.S. Nuclear Regulatory Commission. Doses in Our Daily Lives (available at https://www.nrc.gov/about-nrc/radiation/around-us/doses-daily-lives.html).

[12] National Aeronautics and Space Administration. Solar Storm's Radiation at Martian Orbit and Surface (available at https://mars.nasa.gov/resources/solar-storms-radiation-at-martian-orbit-and-surface/).

[13] Gebhardt, Chris. (December 28, 2017). "Year in Review, 2017 (Part II): Rovers, Orbiters Peel Away Mars' Secrets & Reveal New Mysteries." Spaceflight.com (available at https://www.nasaspaceflight.com/2017/12/yir-2017-part-ii-rovers-mars-secrets-reveal-mysteries/).

[14] Zeitlin, C. et al. (2014). National Aeronautics and Space Administration, NASA Human Research Program Investigators' Workshop (available at https://three.jsc.nasa.gov/iws/25/physicsSpaceTech/3314.pdf).

[15] Seedhouse, E. (2018). Acute Radiation Sickness. In: Space Radiation and Astronaut Safety. Springer Briefs in Space Development. Springer: Cham, Switzerland.

[16] National Aeronautics and Space Administration. Solar Minimum is Coming (available at https://science.nasa.gov/science-news/news-articles/solar-minimum-is-coming).

[17] Michalski, Joseph R.; Bleacher, Jacob E. (2013). "Supervolcanoes Within an Ancient Volcanic Province in Arabia Terra, Mars." Nature (available at http://www.nature.com/nature/journal/v502/n7469/full/nature12482.html).

[18] European Space Agency. (March 12, 2015). The Ages of Mars (available at http://sci.esa.int/mars-express/55481-the-ages-of-mars/).

[19] Carr, M. H. (2007). "Mars: Surface and Interior in Encyclopedia of the Solar System," McFadden, L.-A. et al., Eds. Academic Press: San Diego, CA, p. 321.

---

[20] Short, N. M. Volcanic Landforms and Surface Features: A Photographic Atlas and Glossary. New York: Springer-Verlag, pp. 1–18.

[21] Meresse, S; Costard, F; Mangold, N.; Masson, Philippe; Neukum, Gerhard; the HRSC Co-I Team. (2008). "Formation and Evolution of the Chaotic Terrains by Subsidence and Magmatism, Hydraotes Chaos, Mars." Icarus. 194 (2): 487.

[22] Pinkerton, H.; Bagdassarov, N. "Transient Phenomena in Vesicular Lava Flows Based on Laboratory Experiments with Analogue Materials." Journal of Volcanology and Geothermal Research (available at http://www.sciencedirect.com/science/article/pii/S037702730300341X).

[23] Pinkerton, H. (1996). "Rheological Properties of Basaltic Lavas at Sub-Liquidus Temperatures." Journal of Volcanology and Geothermal Research (available at http://cat.inist.fr/?aModele=afficheN&cpsidt=5970696).

[24] Léveillé, Richard J. ; Datta, Saugata. (2010). "Lava Tubes and Basaltic Caves as Astrobiological Targets on Earth and Mars: A Review." Planetary and Space Science. 58 (4): 592–598.

[25] Underground Towns on the Moon and Mars: Future Human Habitats Could Be Hidden in Lava Tubes. Spaceflight Insider (available at http://www.spaceflightinsider.com). Archived from the original on December 1, 2017. Retrieved May 4, 2018.

[26] Cushing, Glen E. et al. (2014). "Atypical Pit Craters on Mars: New Insights from THEMIS, CTX, and HiRISE observations". Journal of Geophysical Research: Planets (available at https://agupubs.onlinelibrary.wiley.com/doi/pdf/10.1002/2014JE004735).

[27] Cushing, Glen E. et al. (2014). "Atypical Pit Craters on Mars: New Insights from THEMIS, CTX, and HiRISE observations". Journal of Geophysical Research: Planets (available at https://agupubs.onlinelibrary.wiley.com/doi/pdf/10.1002/2014JE004735).

[28] Volcanic and Geologic Terms (available at http://volcano.und.edu). Archived from the original on May 14, 2008.

[29] Distribution, Morphology, and Origins of Martian Pit Crater Chains (available at http://www.agu.org).

[30] National Aeronautics and Space Administration. Pit Crater near Elysium Mons (available at https://www.jpl.nasa.gov/spaceimages/details.php?id=PIA20368).

[31] Léveillé, Richard J.; Datta, Saugata. (2010). "Lava Tubes and Basaltic Caves as Astrobiological Targets on Earth and Mars: A Review." Planetary and Space Science. 58 (4): 592–598.

[32] Walden, Bryce; Billings, Thomas; York, Cheryl; Gillett, Stephen; Herbert, Mark. "Utility of Lava Tubes on Other Worlds." Planetary Society. Archived from the original on April 13, 2014. Retrieved February 25, 2014.

[33] Dvorsky, George (October 18, 2014), "Could This Lunar Cave Provide Shelter for a Future Moon Colony?", io9/ Gizmodo.com, retrieved 24 January 2016.

[34] Coombs, Cassandra R.; Hawke, B. Ray (September 1992), "A search for intact lava tubes on the Moon: Possible lunar base habitats", In NASA. Johnson Space Center, The Second Conference on Lunar Bases and Space Activities of the 21st Century (SEE N93-17414 05-91), 1, pp. 219–229.

[35] Cain, Fraser. (2018). Living Underground on Other Worlds. Exploring Lava Tubes. Universe Today. (available at https://www.universetoday.com/139021/living_underground_exploring_lava_tubes/)

[36] Hellas Planitia. Gazetteer of Planetary Nomenclature. USGS Astrogeology Science Center (available at https://web.archive.org/web/20161225210458/https://astrogeology.usgs.gov/search/map/Docs/Globes/i2782_sh1).

[37] Schultz, Richard A.; Frey, Herbert V. (1990). "A New Survey of Multi-Ring Impact Basins on Mars." Journal of Geophysical Research. 95: 14175

[38] National Aeronautics and Space Administration. Hellas Planitia (available at https://mars.nasa.gov/resources/7648/topographic-map-of-hellas-planitia/?site=insight).

[39] Haberle, Robert M. et al. (October 25, 2001). "On the Possibility of Liquid Water on Present-Day Mars." Journal of Geophysical Research. 106 (E10): 23,317–23,326.

[40] Fairén, Alberto; Dohm, James; Uceda, Esther; Rodríguez, Alexis; Fernández-Remolar, David; Schulze-Makuch, Dirk; Amils, Ricardo. "Prime Candidate Sites for Astrobiological Exploration Through the Hydrogeological History of Mars." Planetary and Space Science.

[41] Calderón, L.; Robertson, K.; Tovar, D. (2015). Geomorphologic Evolution of the Zone of Hadriacus Patera in Mars. 46th Lunar and Planetary Science Conference, 2014.

[42] Calderón, L.; Robertson, K.; Tovar, D. (2015). Geomorphologic Evolution of the Zone of Hadriacus Patera in Mars. 46th Lunar and Planetary Science Conference, 2014.

[43] "Underground Towns on the Moon and Mars: Future Human Habitats Could Be Hidden in Lava Tubes." Spaceflight Insider (available at http://www.spaceflightinsider.com). Archived from the original on December 1, 2017. Retrieved May 4, 2018.

[44] Nobel, Park S. (2009). Solar Radiation. Physicochemical and Environmental Plant Physiology (4th Edition) (available at https://www.sciencedirect.com/topics/biochemistry-genetics-and-molecular-biology/solar-radiation).